\documentclass[aps,showpacs,preprintnumbers,amsmath,amssymb, 
twocolumn, tightenlines]{revtex4}

\usepackage[dvips]{graphicx}\usepackage{bm}
\usepackage{amsmath}
\usepackage{dcolumn}
\usepackage{bm}

\newcommand{\be}{\begin{eqnarray}}
\newcommand{\ee}{\end{eqnarray}}

\begin{document}
\title{How much entropy is produced in strongly coupled \\
Quark-Gluon Plasma (sQGP) by dissipative effects?}

\author{M.Lublinsky and E.Shuryak}
\affiliation{ 
Department of Physics and Astronomy, State University of New York, 
Stony Brook NY 11794-3800, USA
}

\date{\today}
\begin{abstract}
We argue that estimates of dissipative effects based on 
the first-order hydrodynamics with shear viscosity are 
 potentially  misleading because
 higher order terms in the gradient expansion of the dissipative part
of the stress tensor tend to reduce them. 
Using recently obtained sound dispersion relation in thermal $\cal
N$=4
supersymmetric plasma, we calculate the $resummed$ effect of
these high order terms for Bjorken expansion appropriate
to RHIC/LHC collisions. A reduction of entropy production
is found to be substantial, up to an order of magnitude. 
\end{abstract}

\vspace{0.1in}

\maketitle

  Hydrodynamical description of matter created in high energy
  collisions have been proposed by Landau \cite{Landau} more than 50 years
 ago, motivated by large coupling at small distance, as followed
from the beta functions of QED and scalar theories known at the time.
Hadronic matter is of course described by QCD, in which the
coupling runs in the opposite way. And yet, recent RHIC experiments
have shown spectacular collective flows, well described by relativistic
hydrodynamics. More specifically, one observed three types of flow:
(i)  outward expansion in
transverse plane, or radial flow,
 (ii) azimuthal asymmetry or ``elliptic flow'' 
\cite{Teaney:2000cw,Kolb:2003dz},
as well as recently proposed (iii) ``conical flow'' from quenched jets
\cite{Casalderrey-Solana:2004qm}. These observation lead to conclusion that 
QGP at RHIC is a near-perfect liquid, in a strongly coupled
regime \cite{Shu_liquid}.
  The issue we discuss below is at what ``initial time''  $\tau_0$
one is able to start hydrodynamical description of heavy ion
collisions,
without phenomenological/theoretical contradictions.

 Phenomenologically, it was argued in 
\cite{Teaney:2000cw,Kolb:2003dz}
 that elliptic flow is especially sensitive to $\tau_0$. 
 Indeed,  ballistic
motion of partons may quickly erase the initial spatial anisotropy
on which this effect is based. In practice, hydrodynamics at RHIC is
usually used starting from 
 time $\tau_0\sim 1/2 fm$, otherwise the observed 
ellipticity is not reproduced.

Can one actually use hydrodynamics  reliably at such short
 time? How large is $\tau_0$   compared to a relevant
``microscopic scales'' of sQGP?  How much dissipation occurs
in the system at this time?  As a measure of that, we will 
calculate below  the ratio of the amount of entropy produced
at $\tau>\tau_0$ to its ``primordial'' value at $\tau_0$,
$\Delta S/S_0$.

  To set up the problem, let us start with a very crude dimensional
 estimate. If we think that the QCD effective
coupling is large $\alpha_s\sim 1$ and the only reasonable microscopic
length is given by  temperature \footnote{Note we have ignored
  e.g. $\Lambda_{QCD}$.}, then the relevant micro-to-macro
ratio of scales is simply
$T_0\tau_0$. With $T_0\sim 400 \, MeV$  at RHIC, one finds 
this ratio to be
 close to one.  We are then lead to a pessimistic
conclusion: at such time application of any macroscopic theory, thermo- or
 hydro-dynamics,  seems to be impossible, since order one 
corrections are expected.  
  
   Let us then do the first approximation, including
the explicit viscosity term to the first order.
%
Zeroth order (in mean free path) stress  tensor 
used in the ideal hydrodynamics  has the form
\be
T_{\mu\nu}^{(0)}= (\epsilon+p)\,u_\mu u_\nu\,+\,p\,g_{\mu\nu}
\ee
while dissipative corrections are induced by gradients 
 of the velocity field. The well known
 first order corrections are due to 
shear ($\eta$) and bulk ($\xi$) viscosities
 \be \delta T_{\mu\nu}^{(1)}=\eta(\nabla_\mu u_\nu + \nabla_\nu u_\mu
 -{2\over 3}\Delta_{\mu\nu}\nabla_\rho u_\rho)+\xi(\Delta_{\mu\nu}\nabla_\rho u_\rho) 
  \ee
 In  this  equation  the  following  projection
 operator onto the matter rest frame was used:
  \be \nabla_\mu\equiv\Delta_{\mu\nu}\partial_\nu, \,\,\,\ \ 
  \Delta_{\mu\nu}\equiv g_{\mu\nu}-u_\mu u_\nu \ee
The energy-momentum conservation $\partial^\mu\,T_{\mu\nu}$ 
at this order corresponds to Navier-Stokes equation.

  Because
colliding nuclei are Lorentz-compressed, 
the largest gradients at early time are longitudinal, along the
beam direction. The expansion at this time can be approximated 
by well known
 Bjorken rapidity-independent setup \cite{Bjorken},
in which hydrodynamical equations depend on only one coordinate --
proper time $\tau=\sqrt{t^2-x^2}$. 
  \be \label{s}
{1 \over\epsilon+p} {d\epsilon \over d\tau} ={1\over s}{ds\over d\tau}=-{1 \over
   \tau}\left(1-{(4/3)\eta+\xi \over (\epsilon+p)  \tau}\right)
      \ee
where we have 
introduced the entropy density $s=(\epsilon+p)/T$. Note that for traceless $T_{\mu\nu}$ (conformally 
invariant plasma), the bulk viscosity $\xi=0$. 

  For reasons which will  become clear soon, let us compare this eqn
 to another problem, in which large longitudinal gradients
appear as well, namely sound wave in the medium. 
The dispersion relation (the pole position) for a sound wave with frequency
$\omega$ and  wave vector $q$ is, at small $q$
\be\label{omega}
 \omega=c_s q-{i\over 2} q^2 \Gamma_s, \qquad \Gamma_s\equiv {4\over 3}
{\eta \over \epsilon+p}\ee
  Notice that the right hand side of (\ref{s}) 
contains precisely the same combination of viscosity and thermodynamical
parameters as appears in the sound attenuation problem:
the length $\Gamma_s$, which 
 measures directly  the magnitude of the dissipative 
corrections. At proper times
$\tau\sim\Gamma_s$ one 
has to abandon the hydrodynamics altogether, as the dissipative
   corrections cannot be ignored.

For the entropy production (\ref{s}) the first correction
to the ideal case is  $(1-\Gamma_s/\tau)$. 
Since the correction to one is negative,
 it reduces the rate of the entropy decrease with time.
Equivalently statement is that the total positive sign shows that
some amount of entropy is
generated by the dissipative term. 
 Danielewicz and Gyulassy \cite{DG} have 
analyzed  eq. (\ref{s}) in great details considering  
 various values of $\eta$. Their results indicate
 that the entropy production can be substantial.  


Our present study is motivated by the following argument. If the
hydrodynamical description is forced to begin at early time $\tau_0$ which is $not$ large
compared to the intrinsic micro scale $1/T$, then limiting dissipative effects
to the first gradient only ($\delta T^{(1)}_{\mu\nu}$) is parametrically not 
justified and higher order 
terms have to be accounted for. Ideally those effects need to be 
 $resummed$. As a first step, however, we may attempt to guess their sign and estimate
the magnitude.

 Formally one can think of the dissipative
part of the stress tensor $\delta T_{\mu\nu}$  
 as expended in a series containing all derivatives of the velocity field $u$,
$\delta T_{\mu\nu}^1$ being the first term in the expansion.
In general 3+1 dimensional case there are  many structures, each entering
with a new and independent viscosity coefficient. We call them ``higher order 
viscosities'' and the expansion is somewhat similar to  twist expansion.
For 1+1 Bjorken problem, the appearance of the extra terms modifies eq. 
(\ref{s}), which  can be  written as a series in inverse
proper time
\be\label{s1} {\partial_\tau (s \tau)\over s\,(\tau\,T)}
\,=\,
4\,{\eta\over s}\, \left[{1\over 3}\,{1\over (\tau\,T)^2}\,+\,
\sum_{n=2}^\infty {c_n \over (T \tau)^{2n}}\right]
\ee
We have put $T$ here simply for dimensional reasons: clearly $T\tau$
is a micro-to-macro scale ratio which determines convergence
 of these series and the total amount of produced entropy.
Similarly, the sound wave dispersion relation becomes nonlinear
as we go beyond the lowest order:
\be \label{om} 
&&\omega= \Re [\omega(q)] \ + \ i\,\Im [\omega (q)]\,;\\
&&{\Re [\omega]\over 2\,\pi\,T}
\,=\,c_s\,{q\over 2\,\pi\,T}\ 
+\ \sum_{n=1}^\infty r_n\,\left({q\over 2\,\pi\,T}\right)^{2\,n+1}; \nonumber
\\
&&{\Im[\omega]\over 2\pi T}\,=\,-\,{4\pi\eta\over s}\left[
\,{1\over 3}\,\left({q\over 2\pi T}\right)^2\ +\ 
\sum_{n=2}^\infty \eta_n\,
\left({q\over 2\pi T}\right)^{2\,n}\right] \nonumber
\ee
Based on T-parity arguments we keep only odd (even) powers of $q$ for 
the real (imaginary) parts of $\omega$. The coefficients $c_n$, $r_n$ and
$\eta_n$ are related since they originate from the very same gradient
expansion of $T_{\mu\nu}$. Although both the entropy
production series  above and sound absorption should converge to
sign-definite answer, the coefficients of the series may well be
of alternating sign (as we will see shortly).

Clearly, keeping 
these next order  terms can  be useful only provided there is some
microscopic theory which would make it possible to determine
the values of the high order
viscosities. For strongly coupled QCD plasma this information
is at the moment beyond  current theoretical reach, and we have to rely on models.
A particularly useful and widely studied model of QCD plasma is ${\cal N}=4$ 
supersymmetric plasma, which is also conformal (CFT).  The
AdS/CFT correspondence \cite{Maldacena} (see \cite{Aharony} for review)
relates the strongly coupled gauge theory description 
to weakly coupled gravity problem in the background of AdS$_5$ black hole  metric.
Remarkably, certain information on higher order viscosities in the CFT plasma
can be read of from the literature and we exploit this possibility below.

The viscosity-to-entropy ratio ($\eta/s=1/4\pi$) deduced from  AdS  
\cite{KSS}
turns out to be quite a reasonable approximation to the values appropriate
for the RHIC data description. Thus one may hope that the information on
the higher viscosities gained from the very same model can be well trusted
as a model for QCD. Admittedly having no convincing argument in favor, 
we simply assume that the viscosity expansion of the QCD plasma displays
very similar  behavior, both qualitative and quantitative, as its CFT sister.
  
\begin{figure}[t]
   \includegraphics[width=4.5cm,angle=-90]{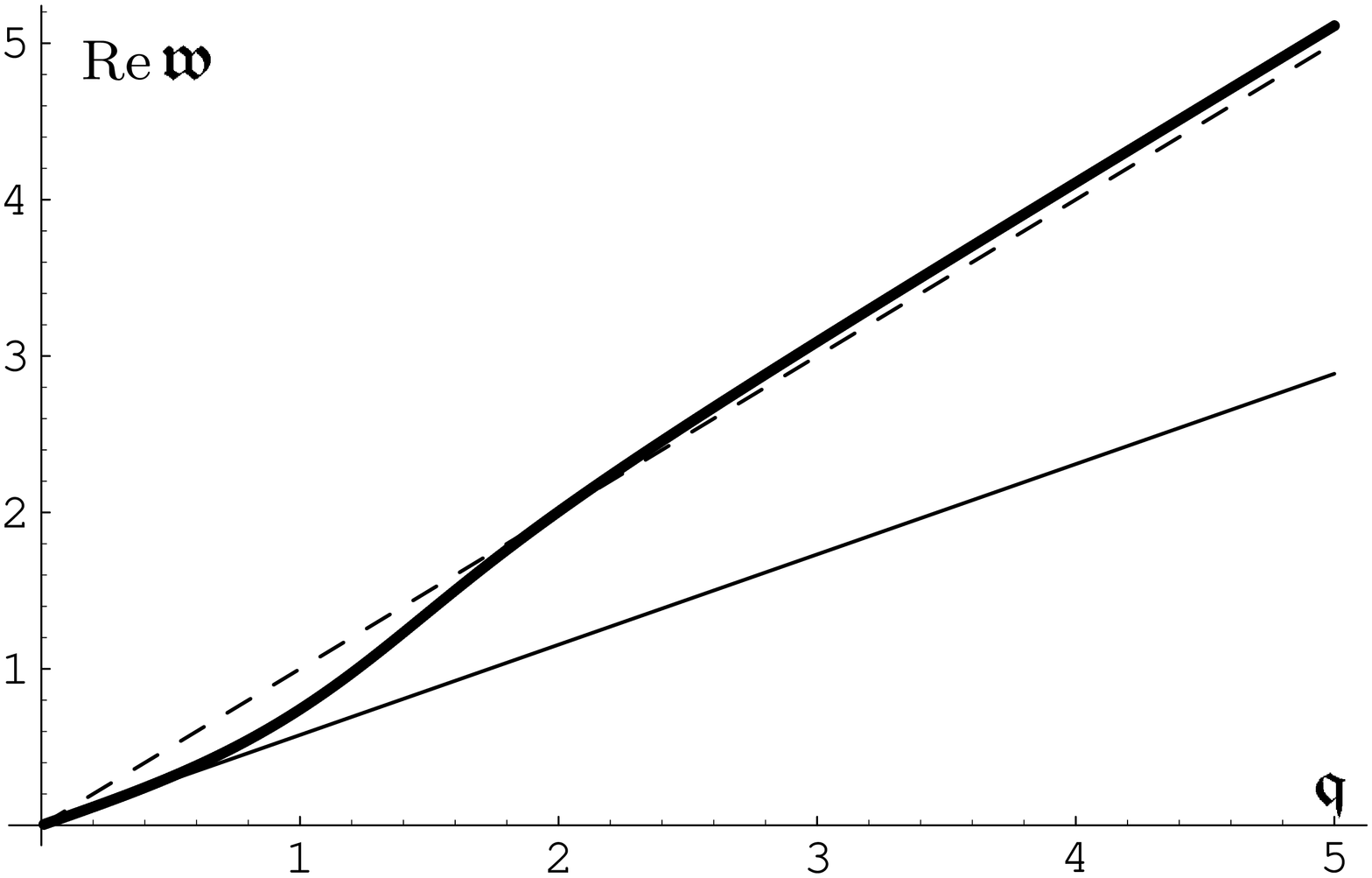} \\
 \includegraphics[width=5cm,angle=-90]{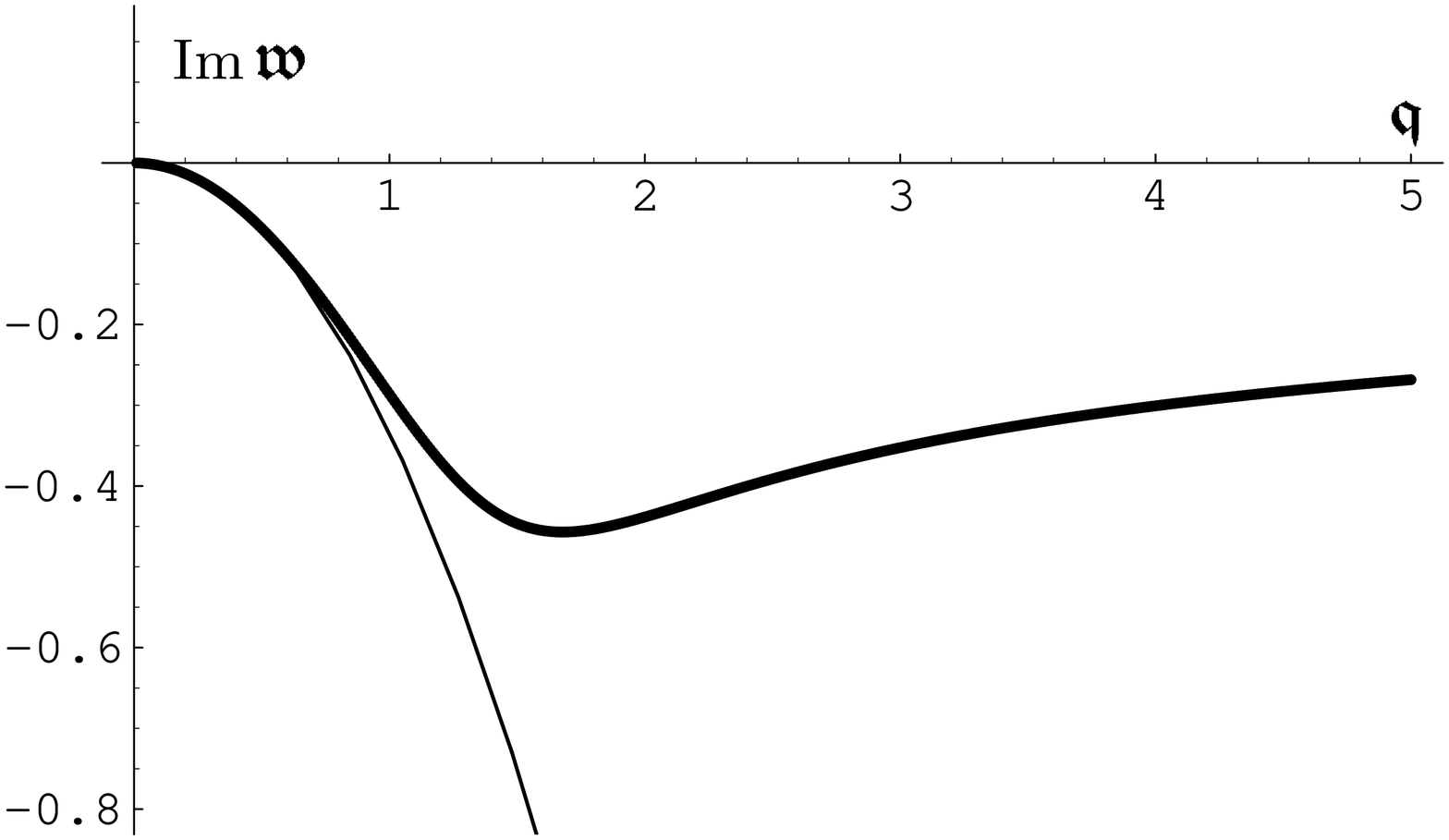}
  \vspace{0.1in}
\caption{Sound dispersion (real and imaginary parts) obtained  from the analysis 
of quasinormal modes in the AdS black hole background. The result and figure are taken 
from  Ref.\protect\cite{KS}.
\label{fig_sound}}
\end{figure}

Our  estimates are  based  on the analysis of the quasinormal modes 
in the AdS black hole background due to Kovtun and Starinets  \cite{KS}.
The dispersion relation for the sound mode, calculated in ref. \cite{KS}, 
is shown in Fig.\ref{fig_sound}. The real and imaginary parts of $\omega$
correspond to the expressions given in (\ref{om}). At $q\rightarrow 0$
they  agree with the leading order hydrodynamical dispersion 
relation (\ref{omega}).

The first important observation is that the next order  coefficient
$\eta_2$ is $negative$, reducing the effect of the first one
when gradients are large. The second is that $|\Im[\omega]|$
has maximum at $q/2\pi T\sim 1$, and at large $q$ the imaginary part
 starts  to decrease. This means that the expansion (\ref{om}) 
has a radius of convergence $q/2\pi T\sim 1$.

In order to estimate the effect of higher viscosities on the entropy production
in the Bjorken setup we first identify $\tau$ in (\ref{s1}) with $2\pi/q$ in
(\ref{om}). Second  we identify the coefficients $c_n$ with $\eta_n$. 
Both sound attenuation and entropy production in question are one dimensional 
problems  associated with the same longitudinal gradients and
presumably the same physics.
  In practice we use the curve 
for the imaginary part of $\omega$  (Fig. \ref{fig_sound}) as an input for
the right hand side of (\ref{s1}).

\begin{figure}[ht]
   \includegraphics[width=8.5cm]{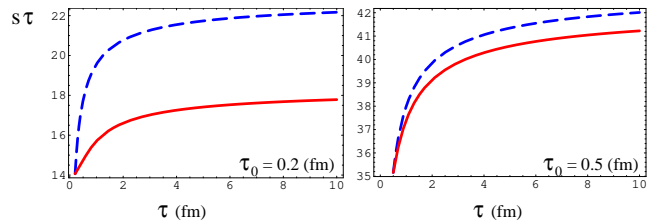}
  \vspace{0.1in}
\caption{Entropy production as a function of proper 
time for initial time $\tau_0=0.2\,{\rm fm}$ (left) and  $\tau_0=0.5\,{\rm fm}$
(right). The initial temperature $T_0=300\,{\rm MeV}$. The dashed (blue) curves
correspond to the first order
(shear) viscosity approximation Eq.(\ref{s}). The solid curve (red)
is the all order dissipative resummation Eq.(\ref{s1}).
\label{res1}}
\end{figure}

\begin{figure}[t]
 \vspace{0.1in}
   \includegraphics[width=8.5cm]{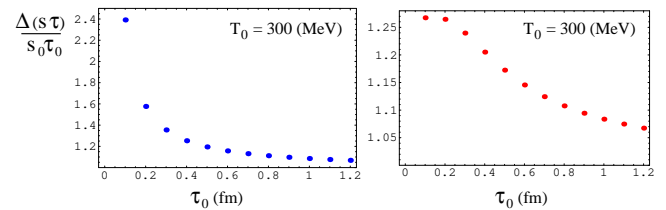} 
\caption{Fraction of entropy produced during the hydro phase as a function
of initial proper time. 
The initial temperature $T_0=300\,{\rm MeV}$. 
The left (blue) points
correspond to the first order
 (shear) viscosity approximation. The right (red) points are 
for the all order  resummation.
\label{res2}}
\end{figure}

The numerical results are shown in Figs. \ref{res1} and \ref{res2} in which
we compare our estimates with the ``conventional'' shear viscosity results
from (\ref{s}). To be fully consistent with the model we set 
$\eta/ s=1/4\pi$. We also set the initial temperature 
$T_0=300\,{\rm MeV}$ while the standard equation of state $s\,=\,
4 \,k_{SB}\, T^3$. For the coefficient  $k_{SB}$ we use the ``QCD'' value
$$k_{SB}= {\pi^2\over 90}\left( 2(N_c-1)^2\, +\, {7\over 2}N_c n_f\right);~  ~
~ ~ ~ n_f=3; ~  ~
~ ~ ~N_c=3$$   

Fig. \ref{res1} presents the results for 
entropy production as a function of proper time for two
 initial times $\tau_0=0.2\,{\rm fm}$  and  $\tau_0=0.5\,{\rm fm}$. The dashed
lines correspond to the first order result (\ref{s}) while the solid curves
include the higher order viscosity corrections. Noticeably there is a dramatic
effect toward reduction of the entropy production as we start the hydro 
evolution at earlier times (the effect is almost invisible on the temperature 
profile). This is the central message of the present paper.

Fig. \ref{res2} illustrates the relative amount of entropy produced during 
the hydro phase as a function of initial time. If the fist order hydrodynamics
is launched at very early times, the hydro phase produces too large amount
of entropy, up to 250\%. (Such a large discrepancy 
is not seen in the RHIC data.)  In sharp
contrast, the results from the resummed viscous hydrodynamics is very stable, and does not 
produce more than some 25\% of initial entropy, even if pushed to start from
extremely early times. The right figure displays
the absence of any pathological explosion  at small $\tau_0$. 

It is worth commenting that we carried the analysis using the minimal
value for the 
ratio  $\eta/s=1/4\pi$. We expect that if this ratio is taken larger,
the discrepancy between the first order dissipative hydro and all orders 
will be even  stronger.

Before concluding this paper we note that 
a practical implementation of relativistic viscous
hydrodynamics had followed Israel-Stewart 
second order formalism (for recent publications see 
 \cite{Baier}) in which one introduces additional parameter , the
relaxation time for the system. Then the dissipative part of the stress tensor
is found as a solution of an evolution equation, with the relaxation time
being its parameter. For the Bjorken setup, the dissipative tensor thus 
obtained has all powers in $1/\tau$ and might resemble the expansion 
in (\ref{s1}) and (\ref{om}). The use of AdS/CFT may shed light
on the interrelation between the two approaches: 
the first step in this direction
has been made recently \cite{janik}, resulting in numerically
 very  small relaxation time.

  Finally, why can it be that macroscopic approaches like hydrodynamics
can be rather accurate at such a short time scale? Trying to answer
this central question one should keep in mind that $1/T$ is $not$
the shortest microscopic scale. The inter-parton distance
 is much smaller,
$\sim 1/(T*N_{dof}^{1/3})$ where the number of effective degrees of
 freedom $N_{dof}\sim 40$ in QCD while $N_{dof}\sim N_c^2 \rightarrow\infty$
in the AdS/CFT approach.

In summary, we have argued that the higher order dissipative terms
strongly reduce the effect of the usual viscosity. 
Therefore
an ``effective''  
viscosity-to-entropy ratio found from comparison Navier-Stokes
results to experiment, can  even be below the (proposed)
lower bound of $1/4\pi$. 
We conclude that it is not impossible
 to use a hydrodynamic description of RHIC collision starting from 
very early times. In particular, our study suggests
that the final entropy observed and
its ``primordial'' value obtained right after collision should
indeed match, with an accuracy of 10-20 percent.

\section*{Acknowledgment}

We are thankful to Adrian Dumitru whose 
results (presented in his talk at Stony Brook)
 inspired us to think about the issue of entropy production during the 
hydro phase. He emphasized to us the  important problem
of matching the final entropy measured 
after late hydro stage  with the early-time partonic predictions,
based on approaches such as color glass condensate.
This work is  supported by the US-DOE grants DE-FG02-88ER40388
and DE-FG03-97ER4014.

\end{document}